# Bridging Metal Additive Manufacturing and RF Accelerator Design: Development of a 704.4 MHz Crossbar H-Mode Linac for Efficient Beam Acceleration

Chuan Zhang [1, 2, 3, *], Eduard Boos [1, 2], Roland Böhm [1], Ramy Cherif [1], Alexander Japs [1], Stefan Wunderlich [1, 2]

[1] GSI Helmholtzzentrum für Schwerionenforschung, Planckstr. 1, 64291 Darmstadt, Germany
[2] Institute for Applied Physics, Goethe-University Frankfurt, Max-von-Laue-Str. 1, 60438 Frankfurt am Main, Germany
[3] Helmholtz Research Academy Hesse for FAIR, Max-von-Laue-Str. 12, 60438 Frankfurt am Main, Germany

**ABSTRACT**. The development of Ultra-High Frequency (UHF) linear accelerators via Metal Additive Manufacturing (MAM) is a strategic research focus of the RACERS team at GSI. The 704.4 MHz Crossbar H-mode (CH) cavity, proposed in 2021 to facilitate efficient frequency jumps and downsize accelerator footprints, represents both the highest-frequency CH structure to date and the first of its kind fabricated entirely through MAM. This study demonstrates the structure's capability for efficient beam acceleration in both Continuous Wave (CW) applications (e.g., accelerator-driven systems) and pulsed operations (e.g., spallation neutron sources). By operating in the UHF regime, the cavity inherently enhances sparking resistance, shifting the physical bottleneck away from surface electric field constraints to enable higher accelerating gradients. To manage the resulting thermal loads within compact dimensions, this study utilizes the design freedom of MAM to integrate a sophisticated "lotus-root-like" cooling network, which is a geometry unachievable through conventional subtractive machining. In combination with a kind of high-strength and high-conductivity alloy, CuCr1Zr, the cavity can achieve energy gain rates of 1.4–1.5 MeV/m (CW) and 4.6–4.8 MeV/m (pulsed), while maintaining a peak surface temperature of approximately 60°C. These results indicate that bridging additive manufacturing with advanced RF design provides a robust framework for next-generation UHF linac structures to go beyond current accelerating-gradient limits.

## I. INTRODUCTION

Over the past decades, Crossbar $H_{21(0)}$-mode (CH) cavities have been extensively established as highly efficient structures for proton and ion acceleration in the low-to-intermediate energy range, operating in both room-temperature (RT) and superconducting (SC) regimes [1–3].

Figure 1 depicts the resonant frequencies of the constructed CH cavities as a function of their design beam velocities, where $\beta$ is the beam velocity normalized to the speed of light in vacuum. It highlights a major leap in the highest operating frequency, which almost doubled from 360 MHz for the SC CH prototype (2005) to 704.4 MHz for the RACE (Resonators Additively Constructed for Experiments [4–5]) cavity constructed in 2024. The RACE cavity not only represents the CH-type resonator with the highest resonant frequency achieved to date, but also marks the first realization of such a structure through Metal Additive Manufacturing (MAM).

[*] e-mail: c.zhang@gsi.de; ORCID: 0000-0002-6108-7942

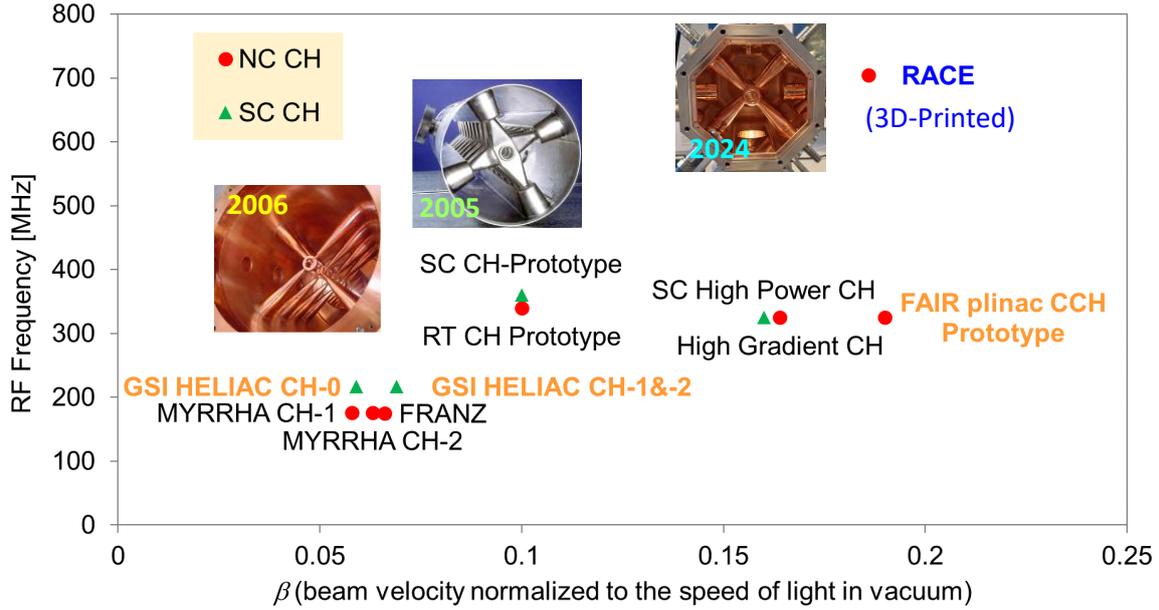

FIG. 1. Overview of the constructed CH cavities. The maximum operating frequency has advanced significantly, from the 360 MHz SC CH prototype (2005) to the state-of-the-art 704.4 MHz RACE cavity (2024). Photos for the RT and SC CH prototypes are sourced from [3].

The application of MAM technologies in particle accelerators was first reported in 2008 as a novel fabrication technique for a copper radio frequency (RF) photoinjector [6]. Its potential for manufacturing SC niobium cavities was subsequently explored in 2015 [7]. At GSI, C. Z. began to conceive and investigate the application of MAM to Ultra-High Frequency (UHF, 0.3–3 GHz) resonant structures in 2017, after taking over the R&D work package for the 1–2 GHz stochastic cooling system of the FAIR Collector Ring [8]. RF structures operating at such high frequencies are characterized by highly compact transverse dimensions. MAM technologies eliminate many constraints of conventional manufacturing processes when fabricating components with complex external geometries or intricate internal structures, making them especially advantageous for realizing compact resonators. However, storage rings typically require an ultra-high vacuum environment. For instance, the CR targets a beam lifetime of 100 s, necessitating a base static pressure of $\leq 3 \times 10^{-9}$ mbar at room temperature without in-situ bakeout [9]. Given these stringent vacuum requirements, it was decided to first implement MAM for UHF linacs as a testbed, where a vacuum level of $10^{-7}$ mbar is generally sufficient.

Taking the Continuous Wave (CW) MYRRHA driver linac [10] as an example, a 704.4 MHz CH structure was investigated to enable an efficient frequency jump at $\beta \approx 0.2$, and then first proposed in 2021 [11–12]. Prior to this proposal, the typical operating frequency for CH cavities had remained below 360 MHz. This high-frequency design aims to considerably downsize the overall footprint and to lower both the construction and operational costs of large accelerators. To reduce the length of a linac, a straightforward way is to increase the resonant frequency, as governed by the fundamental relationship for RF acceleration:

$$L_{\text{cell}} \propto \frac{1}{f} \quad (1)$$

where $L_{\text{cell}}$ is the length of an accelerating cell and $f$ is the resonant frequency. Another advantage of operating at increased frequencies is the capability to sustain stronger peak surface electric fields, $E_{s,\text{max}}$, which allows for higher accelerating gradients, $E_{\text{acc}}$. The relationship between the allowable $E_{s,\text{max}}$ and $f$ was first empirically investigated by Kilpatrick in the 1950s [13]. His findings, characterizing the spark-limited breakdown threshold, can be expressed by the Kilpatrick criterion [14]:

$$f = 1.64 E_k^2 e^{-8.5/E_k} \quad (2)$$

where $f$ is the resonant frequency in MHz and $E_k$ is the Kilpatrick limit of the surface electric field in MV/m. For example, the Kilpatrick limit $E_k$ increases from 18.60 MV/m at 360 MHz to 24.61 MV/m at 704.4 MHz, corresponding to a 1.32-fold improvement in the sparking-limited surface field. It should be noted that the Kilpatrick criterion, established in the last century, is nowadays often considered as a very conservative reference. With modern advancements in surface processing and vacuum technologies, significantly higher surface electric fields are now attainable. For pulsed operations, surface fields up to 2.5 times the Kilpatrick limit (2.5 $E_k$) have been proven feasible, as demonstrated by the operational experience of the CERN RFQ2 accelerator [15]. For CW operations, a Kilpatrick factor, defined as the ratio of $E_{s,\text{max}}$ to $E_k$, of 1.8 was successfully validated by the LEDA

RFQ accelerator [16]. This value has since been widely adopted as a reliable design benchmark for high-power CW linacs, balancing high-gradient performance with operational reliability.

As shown in TABLE I, the 704.4 MHz CH structure proposed in [11–12] features exceptionally compact dimensions: its 22 cm diameter is comparable to that of a football, while its length is only 33.7 cm (approximately 1.5 times a football's diameter). Besides the aforementioned breakdown limits, thermal management poses a critical bottleneck for implementing such a compact structure. High-power-density operations necessitate advanced cooling strategies. Given the stringent reliability requirements for MYRRHA-like accelerators, an intricate and highly efficient water-cooling system is essential for RT accelerators to mitigate thermal expansion and ensure the desired RF field stability. Integrating efficient cooling channels into such a compact and complex geometry renders conventional subtractive machining prohibitively difficult. To realize the 704.4 MHz CH structure, Laser Beam Powder Bed Fusion [17], a primary MAM technique, has been adopted. In this process, a high-power laser selectively melts and solidifies metallic powder point by point and layer by layer, following a sliced 3D CAD model (also known as metal 3D-printing). Compared to conventional manufacturing, MAM offers incomparable design freedom, allowing the integration of intricate, conformal cooling networks that would be otherwise unachievable.

TABLE I. Main design parameters of the 704.4 MHz RACE CH cavity.

| Parameter | Value |
| --- | --- |
| Design particle | proton |
| Frequency $f$ [MHz] | 704.4 |
| Design synchronous energy (for the cavity) $W_s$ [MeV] | 16.6 |
| Design beam velocity $\beta$ | 0.186 |
| Number of accelerating cells $N_{cell}$ | 7 |
| Cavity external length $L_{cavity,external}$ [mm] | 337 |
| Cavity inner radius $R_{cavity,i}$ [mm] | 80 |
| Cavity outer radius $R_{cavity,o}$ [mm] | 110 |
| Drift-tube inner aperture radius $R_{tube,i}$ [mm] | 10 |
| Drift-tube outer aperture radius $R_{tube,o}$ [mm] | 16 |

Like the MYRRHA-injector CH cavities [18], the classical solution to cool a CH electrode is to use telescope-like cooling channels: the stem part of the CH electrode is cooled by a single thick channel that gradually reduces its diameter in several stages, like a telescope, and then splits into two thinner channels in the drift-tube section (see the middle graph in FIG. 2). Limited by conventional manufacturing processes, the cooling channels are typically kept at a relatively large distance from the electrode surface, especially in the stem part. To minimize the temperature rise during operation, particularly in CW operation, one often has to increase the flow velocity of the cooling water.

For water-cooled copper components, maintaining a sufficient service life under corrosive and erosive conditions sets an upper limit on the cooling water velocity. The literature [19] suggests that this limit is approximately 1.2 m/s, beyond which the annual copper loss increases significantly. To prevent water-induced damage, the MYRRHA injector CH cavities employ 316L stainless steel for the electrodes [18], utilizing its superior mechanical hardness compared to copper. However, its thermal conductivity is much lower than that of copper (see TABLE II), so an even higher water flow velocity is required to achieve comparable cooling performance.

Enabled by MAM, the RACE cavity features a brand-new cooling design with lotus-root-like channels [4, 20] (see the right graph in FIG. 2). To maximize cooling efficiency, the design incorporates eight slim curved channels ($\varnothing_{min}$ = 3 mm, see the CT graph in FIG. 2) positioned close to the stem surface. These subsequently converge into four channels traversing the drift-tube section, a complex geometric array otherwise unachievable with conventional manufacturing methods. ANSYS Discovery [21] simulations demonstrate that for the copper electrode shown in FIG. 2 with new lotus-root-like channels, the maximum surface temperature increase is only ~17°C (from 25°C to ~42°C) under a heat load of 600 W with a water flow velocity below 1.2 m/s [4, 5].

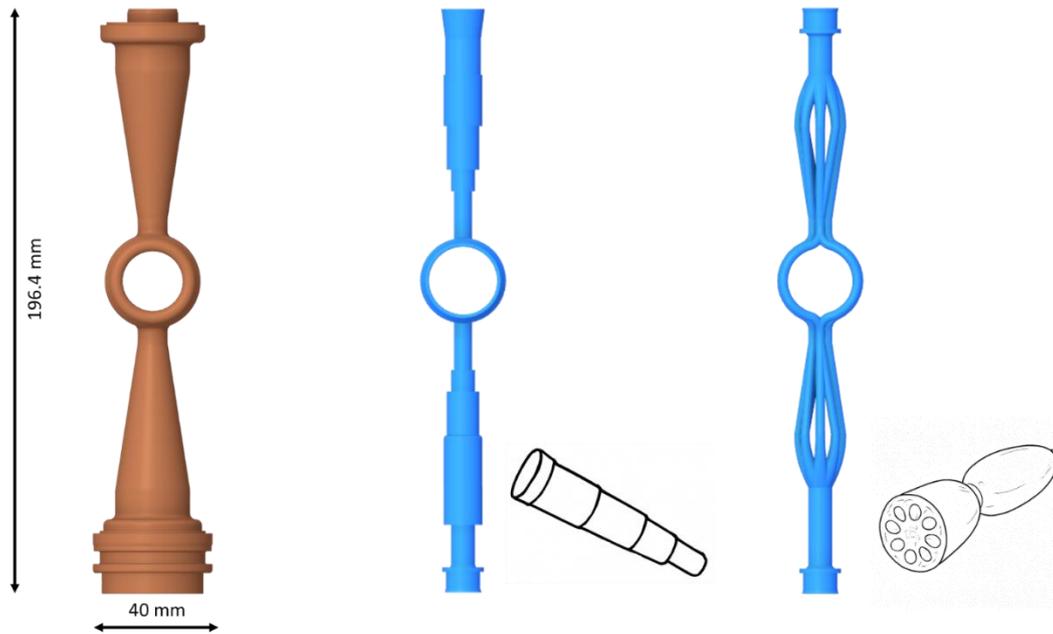

FIG. 2. CH electrode (with dimensions identical to those used for the RACE cavity) and internal cooling designs: electrode appearance (left); classical telescope-like channel (middle); and new lotus-root-like channels enabled by MAM (right). The blue regions represent the internal cooling channels (i.e., the hollow passages within the electrode for water flow).

TABLE II. Properties of construction materials for RT accelerators (copper: typical; 316L stainless steel: adopted for the MYRRHA CH cavities [18]; CuCr1Zr and CuNi2SiCr: recently explored [5]).

| Parameter (at 20°C) | Pure copper | 316L stainless steel | CuNi2SiCr | CuCr1Zr (heat-hardened) |
|---|---|---|---|---|
| Thermal conductivity $\kappa$ [W/(m·K)] | 397 | 15 | 190–240 | 310 |
| Electrical conductivity $\sigma$ [S/m] | $5.8 \times 10^7$ | $1.55 \times 10^6$ | $2.3 \times 10^7$ | $4.3 \times 10^7$ |
| Thermal expansion coefficient $\alpha$ [K$^{-1}$] | $1.65 \times 10^{-5}$ | $1.60 \times 10^{-5}$ | $1.50 \times 10^{-5}$ | $1.70 \times 10^{-5}$ |
| Density $d$ [kg/m³] | 8940 | 7950 | 8840 | 8910 |
| Vickers hardness $HV$ | 45–100 | 150–220 | 160–240 | 110–200 |

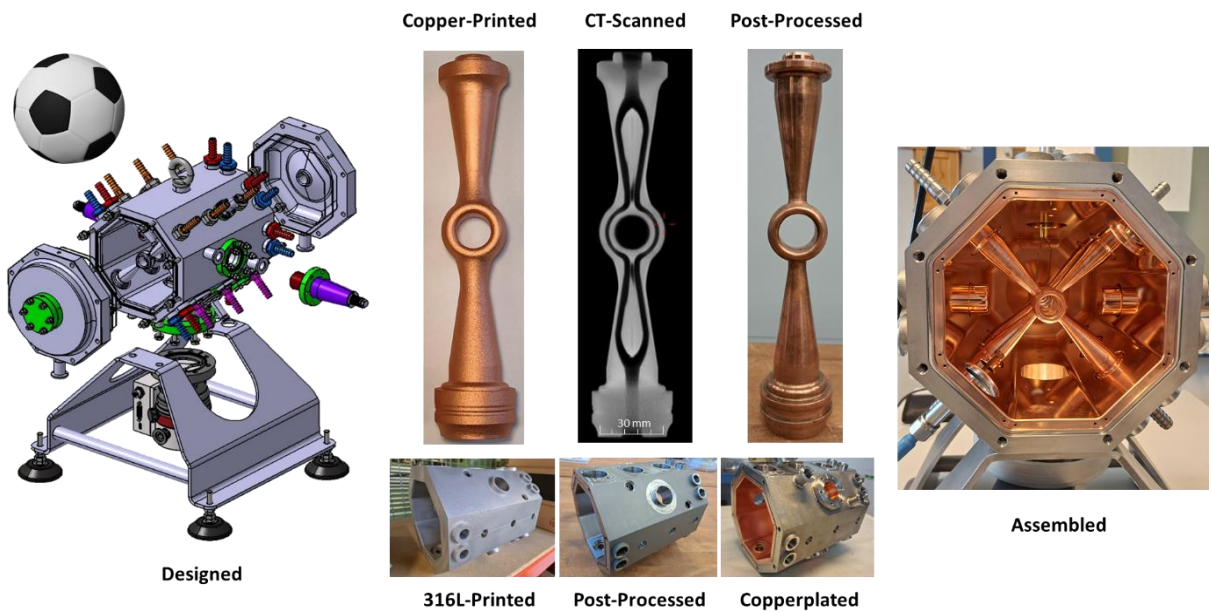

FIG. 3. Development stages of the football-sized RACE cavity: mechanical design and modeling (left); MAM fabrication of the electrodes and the cavity wall (middle); the assembled RACE cavity ready for testing (right).

Based on this unconventional cooling design, the RACE cavity electrodes were printed in pure copper (see FIG. 3). Due to the limited build volume of the adopted printer, the cavity wall was fabricated from 316L stainless steel via MAM. Given that current MAM surface finishes remain insufficient for direct RF applications, a 1 mm machining allowance was added to all functional surfaces (electrodes and cavity wall) during the printing phase. This excess material was subsequently removed via precision mechanical post-processing to achieve the required tolerances. In addition, the 316L-printed cavity wall was copper-plated to ensure the RF surface conductivity required for high-performance linac operation. Detailed evaluations of surface roughness, vacuum compatibility, and low-level RF performance have been reported in [5]. It can be concluded that after proper post-processing, the MAM-fabricated components for the RACE cavity are highly comparable to those produced by conventional subtractive manufacturing.

In [11–12], a 2.7-meter-long beam-transport section comprising four RACE-like CH cavities was proposed to facilitate an efficient frequency jump from 176.1 MHz to 704.4 MHz for 16.6 MeV protons, using the MYRRHA driver linac as an example. Since that study focused on minimizing the frequency-jump-induced longitudinal phase spread (in degrees), the resulting energy gain was of less importance, totaling only 0.4 MeV. Encouraged by the successful realization of the RACE cavity, the present study further investigates the potential of using the 704.4 MHz CH structure, combined with MAM, for high-efficiency beam acceleration, starting from the exit of the aforementioned frequency-jump section.

## II. BEAM DYNAMICS DESIGN

The previous study [11–12] demonstrated the feasibility of a lens-free transition between two 7-gap 704.4 MHz CH cavities for proton beams at $\beta \approx 0.2$. At this beam velocity, space charge effects become less dominant, permitting a highly compact lattice without intermediate transverse focusing. Consequently, the lattice configuration for the present study is optimized to prioritize high-efficiency acceleration: two 14-gap 704.4 MHz CH cavities are employed instead of the previously proposed four 7-gap cavities, with each preceded by a quadrupole triplet to provide the necessary transverse beam focusing.

Like the RACE cavity, each new 14-gap 704.4 MHz cavity is designed to operate at a synchronous phase $\varphi_s = -90°$ with a constant accelerating-cell length $L_{\text{cell}}$, so that each cavity exhibits both transverse and longitudinal symmetry (front and side views), significantly simplifying its mechanical construction and RF tuning. As illustrated in FIG. 4, in the classical beam dynamics design concept, where particles in a beam bunch oscillate about the synchronous particle, the longitudinal acceptance reaches its maximum (360°) when $\varphi_s = -90°$, but no energy gain is obtained for the beam bunch. Therefore, a synchronous phase of about -30° is typically chosen, as it ensures stable longitudinal motion while maintaining relatively high acceleration efficiency. In order to achieve beam acceleration with a $\varphi_s = -90°$ structure, the special EQUUS (EQUidistant mUltigap Structure) beam dynamics method [22] has been adopted. Originally developed for superconducting CH cavities [3, 23], EQUUS utilizes asynchronous beam injection: the phase and energy of the input beam bunch should be less negative and lower than the synchronous phase and energy, respectively. This allows the beam to "slide" through the constant-length cells while gaining energy, a mechanism fundamentally different from the classical design concept. The energy gain within an EQUUS-structured cavity can be estimated by the following expression:

$$\delta W_{\text{cavity}} \approx q V_{\text{eff,cavity}} \cos \overline{\varphi_{\text{bc}}} \qquad (3)$$

where $\delta W_{\text{cavity}}$ denotes the total energy gain of a cavity in MeV, $q$ is the charge state of the particle, $V_{\text{eff,cavity}}$ is the total effective voltage of all cells ("effective" means that the transit-time factor is included) in MV, and $\overline{\varphi_{\text{bc}}}$ is the average bunch-center (bc) phase of all cells.

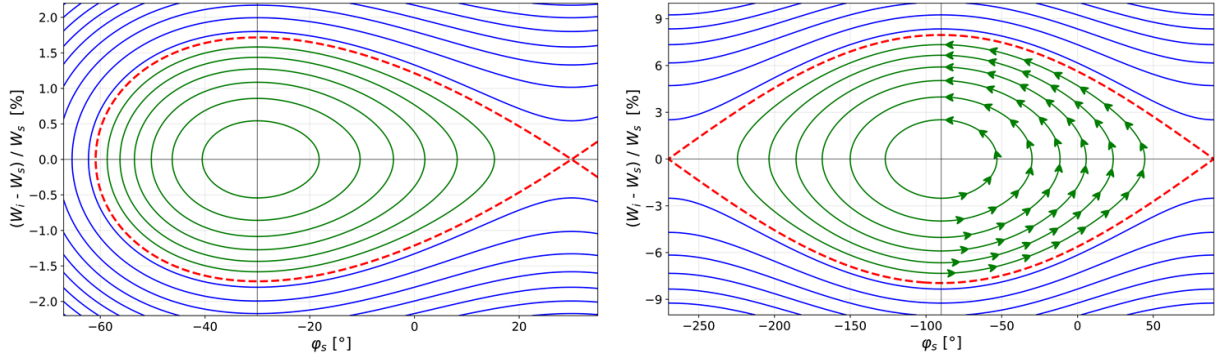

FIG. 4. Schematic comparison of longitudinal phase space trajectories between the classical beam dynamics design concept (left, typically $\varphi_s = -30°$) and the EQUUS method (right, $\varphi_s = -90°$). The red curve represents the separatrix, while the green and blue regions indicate the stable and unstable areas for beam motions, respectively. The green arrows highlight the working area of the EQUUS beam dynamics.

Based on the aforementioned lattice configurations, this study develops two beam dynamics designs for distinct operational requirements. Scenario A (CW mode) is optimized for continuous wave operation required by accelerator-driven systems such as MYRRHA. Scenario B (pulsed mode) targets pulsed operation with duty factors below 10%, typical for spallation neutron sources, e.g., ESS (European Spallation Source, 4%), SNS (Spallation Neutron Source, ~6%), and J-PARC (Japan Proton Accelerator Research Complex, ~1.25%) [24–26], another promising application suggested in [11–12] for the 704.4 MHz CH structure. For the sake of conciseness, they are hereafter referred to as the CW-design and the Pulsed-design, respectively.

For the CW-design, the accelerating gradient of the first 14-gap cavity is set to $E_{acc}$ = 1.76 MV/m, consistent with the highest value among the fifteen 176.1 MHz CH cavities for the MYRRHA injector ($E_{acc}$ = 0.94 – 1.76 MV/m) [27]. Since end-plate thickness does not affect the cavity field (as long as the end plates are not unrealistically thin), the so-called cavity active length defined in Eq. (4) is adopted throughout this study to exclude end-plate thickness influence on all cavity-length-related calculations, e.g., $E_{acc}$ and shunt impedance. This length is calculated as the sum of all individual accelerating-cell lengths (for an EQUUS design, $L_{cell}$ is constant throughout the cavity), i.e.:

$$L_{\text{cavity,active}} = N_{\text{cell}} L_{\text{cell}} = N_{\text{cell}} \beta_s \lambda / 2 \qquad (4)$$

where $L_{\text{cavity,active}}$ is the cavity active length, $N_{\text{cell}}$ is the total number of cells, $\beta_s$ is the normalized synchronous beam velocity, and $\lambda$ is the wavelength in vacuum.

To achieve efficient acceleration and favorable RF and beam performance in both designs, the considerations for selecting the basic parameters, together with the corresponding optimization strategies, are as follows:

- At 704.4 MHz and $\beta_s \approx 0.2$, the cavity active length $L_{\text{cavity,active}}$ is about 60 cm for the 14 cells. Thus, an accelerating gradient $E_{acc}$ = 1.76 MV/m corresponds to an effective cavity voltage $V_{\text{eff,cavity}} \approx 1$ MV. A similar $V_{\text{eff,cavity}}$ is applied to Cavity 2. For the Pulsed-design, each cavity is operated at a $V_{\text{eff,cavity}}$ that is 3.1 times the value of the CW case, corresponding to a duty factor of 9.61% (less than 10%), ensuring that the average RF power loss $P_{\text{loss,cavity}}$ does not exceed the peak power loss of the CW-design.
- To distribute RF power loss more uniformly and reduce the risk of sparking in the new cavities, a flat $V_{\text{eff,gap}}$ profile along the beam axis is adopted, except for the end cells, where amplitude is limited to ~50% by the $H_{21(0)}$ mode.
- The key strategy in designing an EQUUS structure is to properly choose the bunch-center motions. Since $E_{acc}$ has already been determined, the average bunch-center phase can be selected according to Eq. (3) to control the energy gain of each cavity. If $\overline{\varphi_{bc}}$ is closer to -90°, the beam experiences stronger bunching; if it is closer to 0°, more acceleration is obtained. To ensure high beam quality for the designs of this study, the beam trajectory in the longitudinal phase space is optimized for a quasi-symmetry. Specifically, the energy deviation of the bunch center relative to the synchronous energy is kept quite symmetric within each cavity; meanwhile, the initial and final bunch-center phases are kept almost the same so that the beam can experience similar longitudinal focusing conditions at both the entrance and exit of each cavity.

The beam dynamics simulation was performed using the LORASR code [28]. For the CW-design, the output distribution from the 176.1-to-704.4 MHz frequency-jump section [11–12] (containing 98,596 macro-particles) was used as input at 5 mA design beam intensity. Same as in [11–12], the beam dynamics simulations of the 704.4 MHz accelerator employ four times the design beam intensity (i.e., 20 mA) to include the correct charge per bunch for space-charge calculations.

For the Pulsed-design, a particle distribution from the ESS linac at ~17 MeV [29] (containing 97,592 macro-particles) served as the input reference. For a better matching to the lattice optimized primarily for the CW-design, the original ESS distribution was slightly adjusted in the following way:

- The bunch-center energy was shifted from 17.2 MeV to 17.0 MeV.
- Seven macro-particles with energies below 6.5 MeV (extreme outliers from the bunch center) were removed.
- The phase spread (in degrees) was doubled to account for the frequency jump from 352.2 MHz to 704.4 MHz, as this distribution originated from the 352.2 MHz DTL section of the ESS linac.
- While maintaining the same size and shape, the distribution was rotated in both transverse phase spaces to match the defocused state required for the baseline case (i.e., the CW-design).
- The peak design beam intensity in the simulation was doubled from 62.5 mA to 125 mA, also due to the frequency jump.

Figure 5 shows the bunch-center motions in the longitudinal phase space for both the CW-design and the Pulsed-design. Note that design synchronous energies differ between the cavities (see TABLE III and FIG. 6). For the CW-design, the average bunch-center phases $\overline{\varphi_{bc}}$ are -33.6° in both Cavity 1 and Cavity 2, exhibiting good energy gains of 4.6% (Cavity 1) and 4.5% (Cavity 2). In contrast, the Pulsed-design cavities operate with more positive phases with $\overline{\varphi_{bc}}$ = -20.9° (Cavity 1) and $\overline{\varphi_{bc}}$ = -22.6° (Cavity 2), achieving significantly higher energy gains of 15.3% (Cavity 1) and 13.0% (Cavity 2). Figure 6 illustrates the beam-energy evolution for both designs more clearly.

Table III. Beam dynamics design parameters and simulation results of the CW-design and the Pulsed-design.

| Parameter | CW-design (MYRRHA as an example) | | Pulsed-design (ESS as an example) | |
|---|---|---|---|---|
| Design beam intensity [mA] | 5 (20 used in simulation) | | 62.5 (125 used in simulation) | |
| Total number of macro-particles | 98,596 | | 97,592 | |
| Input bunch-center beam energy $W_{bc,i}$ [MeV] | 17 | | 17 | |
| Input emittance $\varepsilon_{x,n.,rms,i}$ [π mm-mrad] | 0.2501 | | 0.3064 | |
| Input emittance $\varepsilon_{y,n.,rms,i}$ [π mm-mrad] | 0.2552 | | 0.2955 | |
| Input emittance $\varepsilon_{z,rms,i}$ [MeV-deg] | 0.1837 | | 0.2926 | |
| Output bunch-center beam energy $W_{bc,o}$ [MeV] | 18.657 | | 22.724 | |
| Emittance growth in $x$ phase space [%] | 0.91 | | 4.03 | |
| Emittance growth in $y$ phase space [%] | 1.37 | | 15.10 | |
| Emittance growth in $z$ phase space [%] | 5.56 | | 13.66 | |
| Beam transmission $T$ [%] | 100 | | 100 | |
| | Cavity 1 | Cavity 2 | Cavity 1 | Cavity 2 |
| Total number of accelerating cells | 14 | 14 | 14 | 14 |
| Effective cavity voltage $V_{eff,cavity}$ [MV] | 0.99 | 1.01 | 3.10 | 3.10 |
| Average bunch-center phase $\overline{\varphi_{bc}}$ [°] | -33.6 | -33.6 | 20.9 | -22.6 |
| Cavity active length $L_{cavity,active}$ [cm] | 56.61 | 57.91 | 58.19 | 62.44 |
| Design synchronous energy $W_s$ [MeV] | 17.41 | 18.25 | 18.43 | 21.31 |

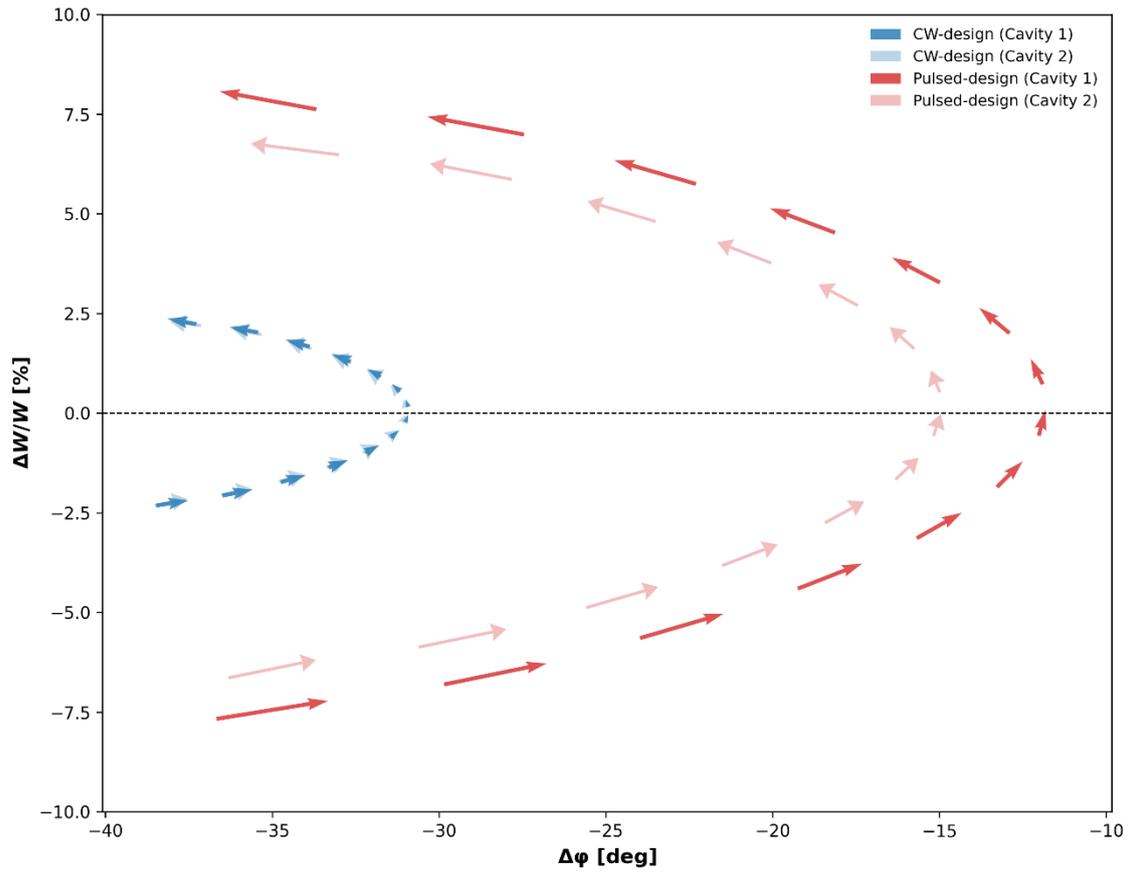

FIG. 5. Bunch-center motions in the longitudinal phase space for both the CW-design and the Pulsed-design. Note that design synchronous energies differ between the cavities (see TABLE III and FIG. 6).

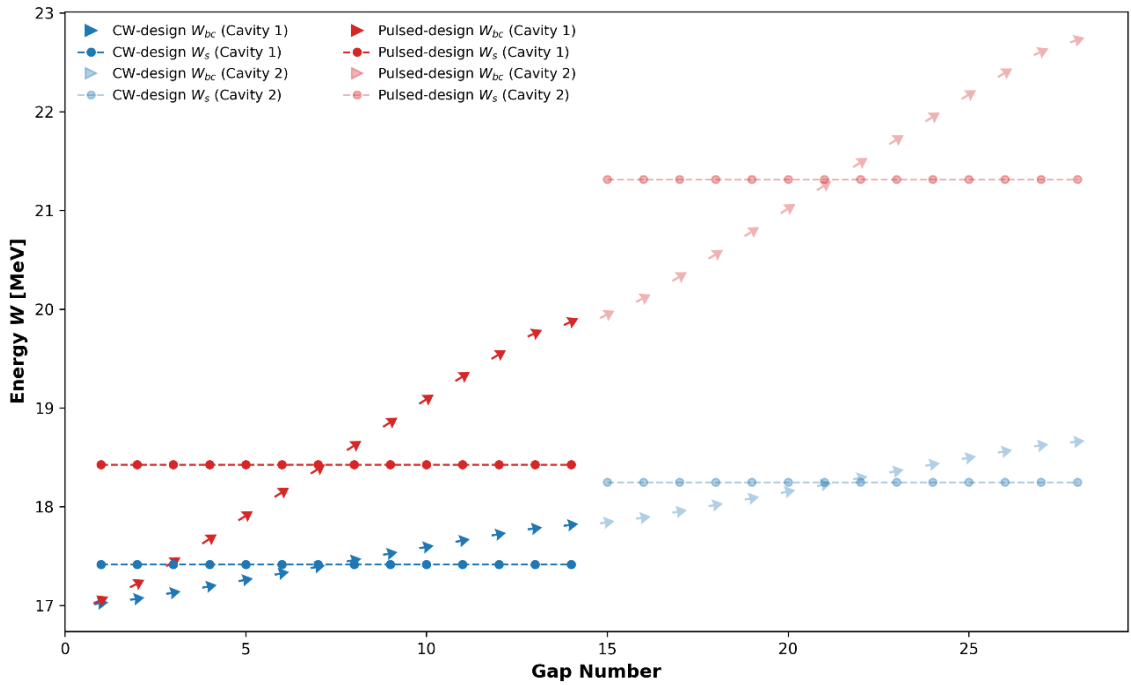

FIG. 6. Evolution of beam energy along the accelerating channels of the CW-design and the Pulsed-design.

Including all particles, the 100% transverse and longitudinal beam envelopes for both designs are shown in FIG. 7. In the transverse planes, the evolution of beam size along the beam axis is similar for both designs, especially in the latter sections, with maximum beam sizes remaining well within the clearance provided by the tube and lens apertures. Meanwhile, the 100% longitudinal beam envelopes indicate that both the relative energy spread and phase spread stay well confined throughout the accelerating channels.

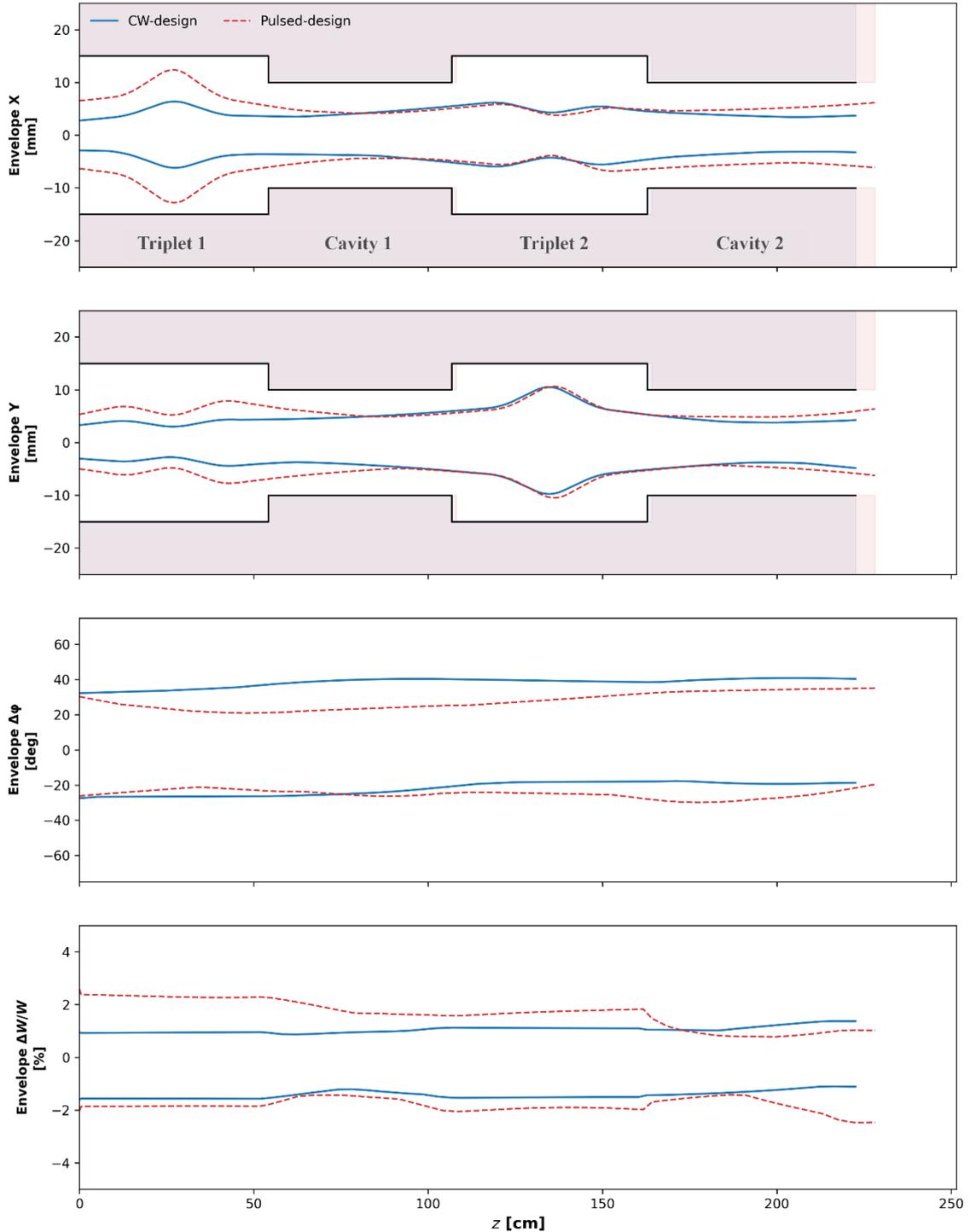

FIG. 7. Comparison of 100% transverse and longitudinal beam envelopes for the CW-design and the Pulsed-design, with cavity and triplet positions indicated.

It should be noted that both designs employ the same two triplets (with Triplet 1 and Triplet 2 having different settings), located at the same positions along the beam line. For the Pulsed-design, which operates at much higher peak beam intensity, the second triplet is set to a correspondingly higher magnetic field in order to counteract stronger space-charge effects, while the maximum pole-tip field remains at only 0.84 T, well within the limits of current quadrupole-magnet technology.

As a function of position along the beam axis, both transverse and longitudinal emittance growths for the two designs are plotted in FIG. 8. In LORASR, emittance values are provided only at the start and end points of the simulation, as well as immediately before and after each lens group. The emittance growths in all phase spaces are modest: ≤ 5.6% for the CW-design and ≤ 15.1% for the Pulsed-design (which operates at significantly higher beam intensity).

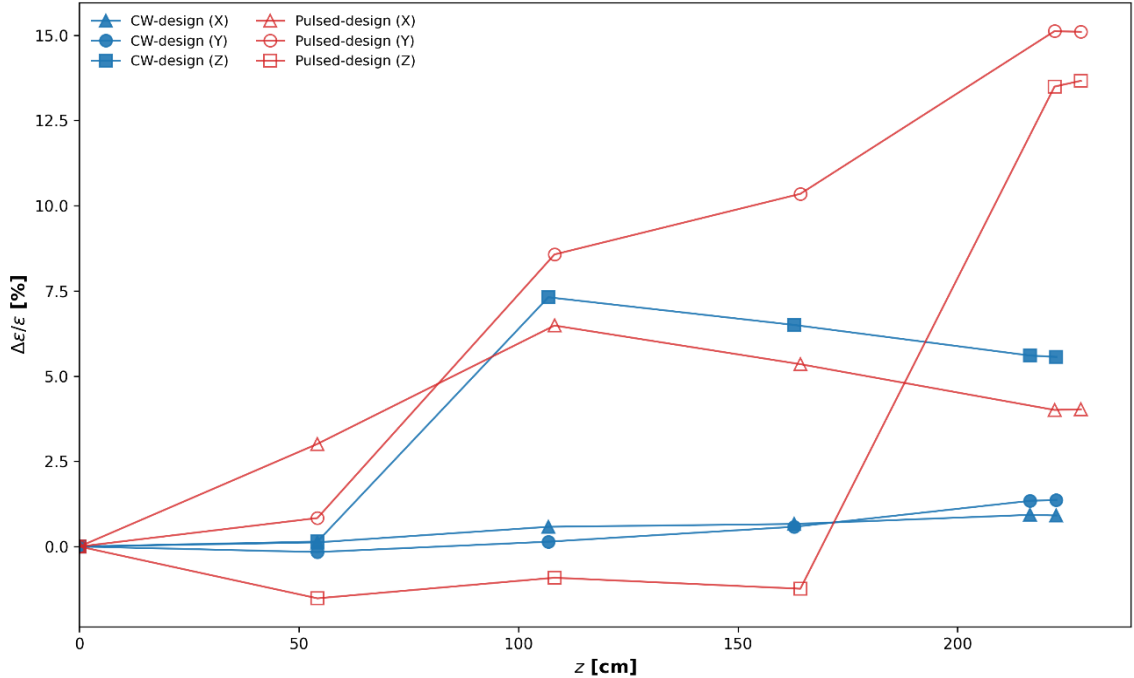

FIG. 8. Evolution of emittance growth in the transverse ($x$, $y$) and longitudinal ($z$) phase spaces along the accelerating channel of the CW-design and the Pulsed-design.

Figure 9 shows the input and output particle distributions for the CW- (top two rows) and Pulsed-designs (bottom two rows), each containing ~100,000 macro-particles. In the transverse planes, both input distributions are defocused (as designed), while the output distributions remain concentrated. In the longitudinal plane, the phase and energy spreads are comparable at the entrance and exit.

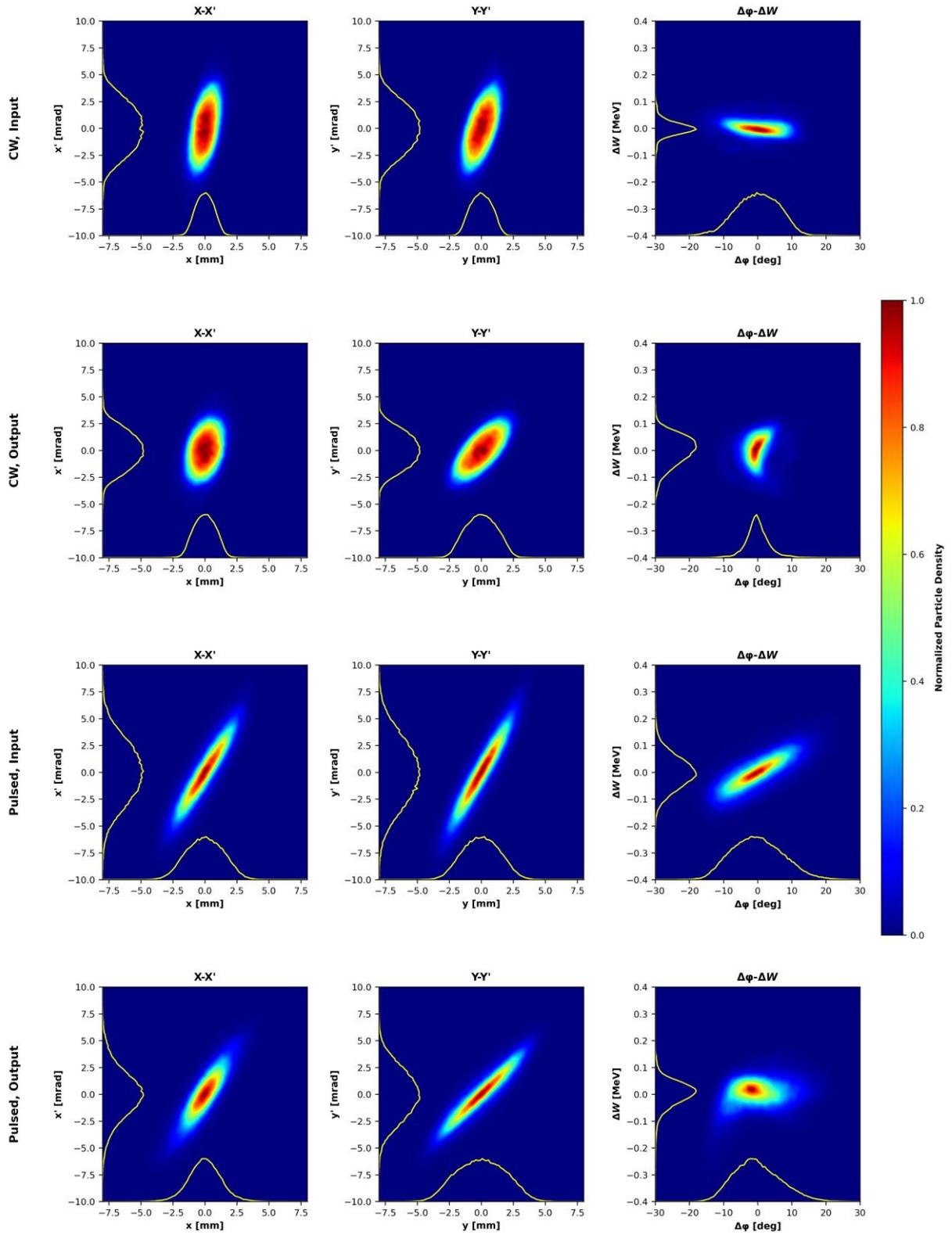

FIG. 9. Input and output particle distributions for the CW-design (top two rows) and the Pulsed-design (bottom two rows).

## III. RF STRUCTURE DESIGN

Following the beam dynamics design, the RF structure design was developed with the help of the CST software [30]. In general, all cavities for both the CW- and Pulsed-designs are essentially identical, except for small differences in accelerating-cell lengths due to different design synchronous energies. This section focuses on the RF structure design of the first CW-design cavity, as the other cavities feature slightly longer cells, which eases the cooling design.

The RACE cavity was designed for low power consumption (< 1.5 kW) and a limited Kilpatrick factor (0.24), so all drift tubes were simply kept identical without voltage-flatness optimization. In contrast, the new 14-gap 704.4 MHz CH cavity is specifically optimized as follows to maximize the accelerating gradient while maintaining operational reliability:

- A cylindrical cavity (in contrast to the octagonal RACE cavity) is adopted to achieve a higher shunt impedance.
- To improve the power-loss distribution and reduce the risk of sparking, the required $V_{eff,gap}$ flatness from the beam-dynamics design is achieved by carefully tuning the drift-tube spacing while keeping the cell lengths (gap-center to gap-center distances) unchanged.

Figure 10 shows the designed 14-gap 704.4 MHz CH cavity, where four tuners in total, including two dynamic tuners, provide a tuning range of -2.727 to +4.071 MHz, sufficient for the expected maximum deviation of ±1 MHz due to mechanical errors or temperature-induced frequency shifts.

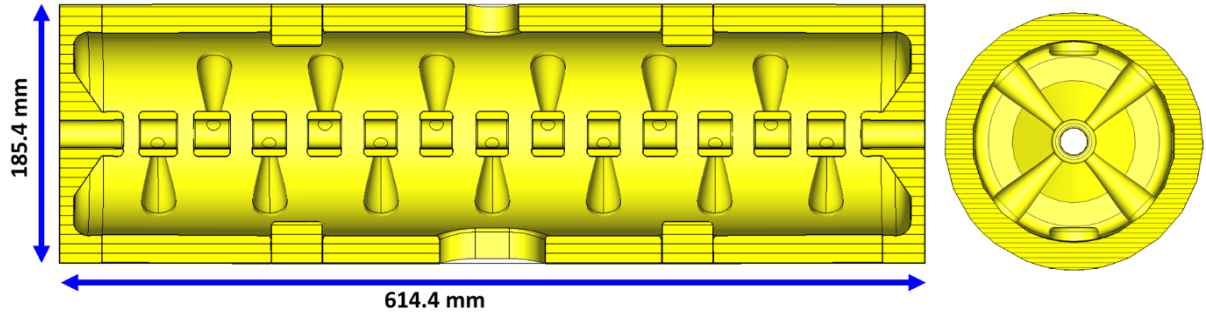

FIG. 10. Cross-sectional views of the CST model of the first 14-gap 704.4 MHz CH cavity of the CW-design.

In FIG. 11, it can be seen that relatively flat on-axis electric-field and $V_{eff,gap}$ distributions along the beam axis are achieved, except for the end cells where the amplitude is limited to ~50% by the $H_{21(0)}$ mode.

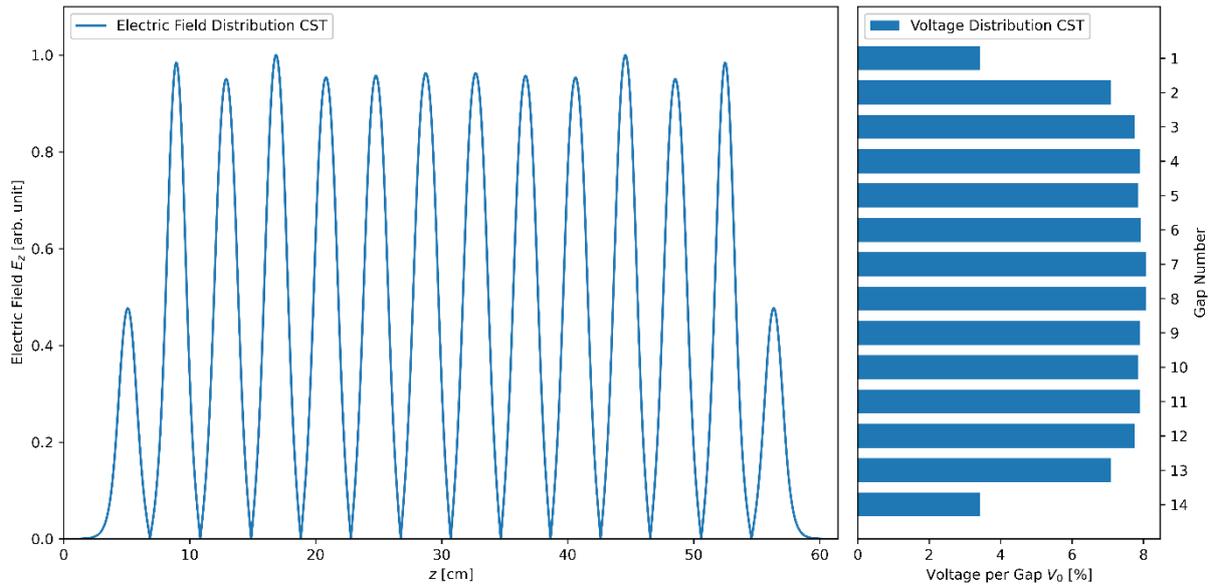

FIG. 11. Simulated on-axis electric-field amplitude and gap-voltage distribution along the first 14-gap 704.4 MHz CH cavity of the CW-design.

The main design parameters are summarized in TABLE IV. They are largely similar to those of the RACE cavity, except for the 7 additional cells (14 vs. 7 cells) and the slightly smaller inner and outer cavity radii.

TABLE IV. Main design parameters of the first 14-gap 704.4 MHz CH cavity of the CW-design.

| Parameter | Value |
|---|---|
| Design particle | proton |
| Frequency $f$ [MHz] | 704.4 |
| Design synchronous energy (for the cavity) $W_s$ [MeV] | 17.4 |
| $\beta$ | 0.190 |
| Number of accelerating cells $N_{cell}$ | 14 |
| Cavity external length $L_{cavity,external}$ [mm] | 614.4 |
| Cavity inner radius $R_{cavity,i}$ [mm] | 72.9 |
| Cavity outer radius $R_{cavity,o}$ [mm] | 92.9 |
| Drift-tube inner aperture radius $R_{tube,i}$ [mm] | 10 |
| Drift-tube outer aperture radius $R_{tube,o}$ [mm] | 16 |

For a copper cavity, the simulated effective shunt impedance is $Z_{eff}$ = 45.1 MΩ/m. In this study, $Z_{eff}$ is defined by:

$$Z_{eff} = \frac{V_{eff,cavity}^2}{P_{loss,cavity} \times L_{cavity,active}} \quad (5)$$

where $Z_{eff}$ is the effective shunt impedance in MΩ/m, $V_{eff,cavity}$ is the total effective voltage of the cavity in MV, $P_{loss,cavity}$ is the RF power loss of the cavity in MW, and $L_{cavity,active}$ is the cavity active length in m.

Using 90% of the simulated $Z_{eff}$ as a safety margin, the calculated RF power loss for the whole cavity is $P_{loss,cavity}$ = 43.34 kW (66.4% on the electrodes, 0.5% on the two lids including the end tubes, and 33.1% on the cavity wall). Figure 12 shows that most electrodes (except the two end ones) exhibit similar surface-current distributions, indicating a relatively homogeneous power-loss distribution. The electrode with maximum $P_{loss,electrode}$ (2.5 kW, 5.768% of $P_{loss,cavity}$), marked by a red frame in FIG. 12, also hosts the location of $E_{s,max}$ = 14.83 MV/m (marked in purple). The corresponding Kilpatrick factor of 0.60 is very safe for CW operation. For the Pulsed-design, with a duty factor of 9.61%, $V_{eff,cavity}$ was increased by a factor of 3.1, yielding a Kilpatrick factor of 1.86 while still remaining fairly safe against sparking, and the corresponding average $P_{loss,electrode}$ on this "hottest" electrode is 2.3 kW, lower than in the CW case.

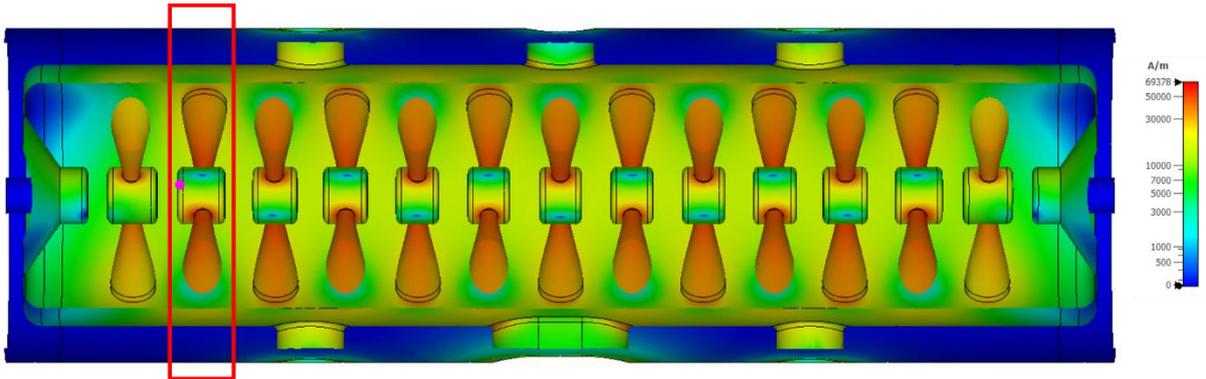

FIG. 12. Simulated surface-current distribution in the first 14-gap 704.4 MHz CH cavity of the CW-design.

## IV. COOLING DESIGN & THERMAL SIMULATION

Based on the RF simulation results of the 14-gap 704.4 MHz CH cavity, a thermal analysis of the "hottest" electrode was performed using ANSYS Discovery. A total heat load of $P_{loss,electrode}$ = 2.5 kW was applied, distributed as follows: 45% to both stems, 35% to the drift tube (restricted to the field region), and the remaining 20% to the transition regions between the stems and the drift tube (see FIG. 13). Due to the modified cavity shape, the electrodes of the 14-gap 704.4 MHz CH cavity are shorter than those of the RACE cavity. However, the section of the electrode inside the cavity, i.e., the field-facing part, remains almost identical to the corresponding part of the RACE electrode, so the same lotus-like cooling channels can still be used, with only the end connection sections shortened.

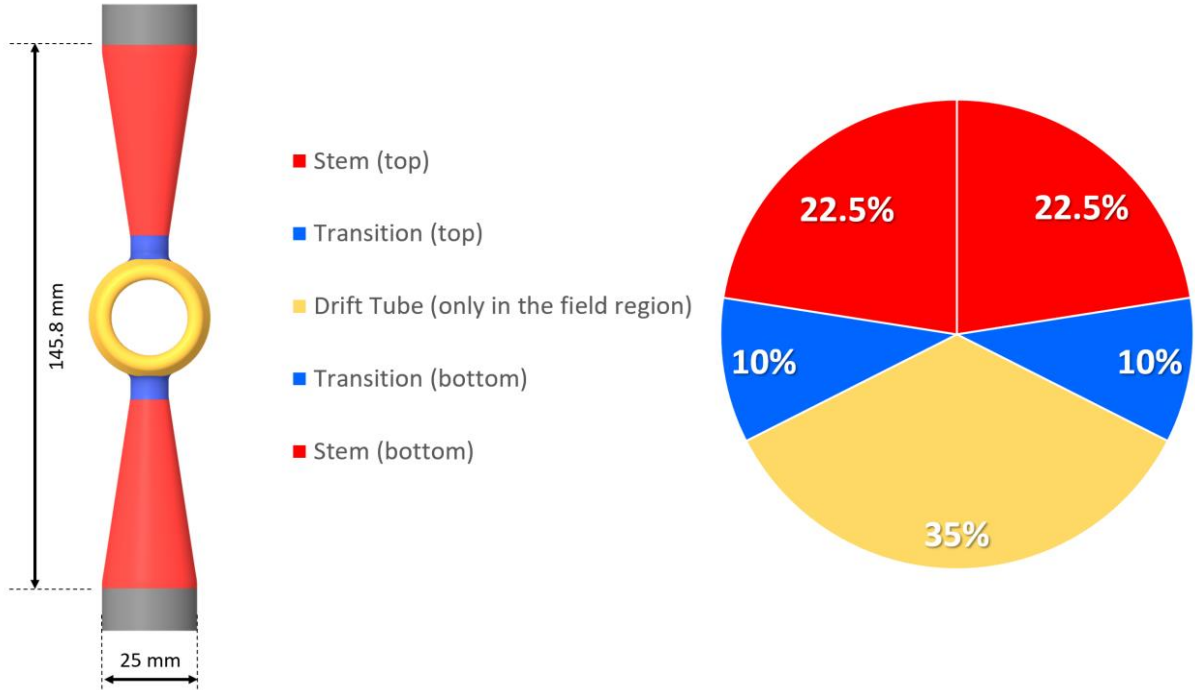

FIG. 13. Applied heat load distribution for the thermal simulation of the 14-gap 704.4 MHz CH cavity electrode (grey-marked regions are embedded within the cavity wall; therefore, no heat load was applied).

Using the thermal simulation of the MYRRHA CH cavities [31] as a reference, the same two boundary conditions were applied to the simulation of the 14-gap 704.4 MHz CH cavity electrode:
- The inlet water temperature, $T_{water,i}$, was set to 17°C.
- The maximum surface temperature on the electrode, $T_{surface,max}$, was limited to ~60°C.

Other adopted simulation conditions are summarized in TABLE V. As shown in Case (a) of FIG. 14, the copper electrode can dissipate a maximum heat load of 1.6 kW, requiring a maximum cooling water velocity of $v_{water,max}$ = 1.22 m/s to maintain $T_{surface,max} \approx 60°C$. To handle the 2.5 kW heat load required by the "hottest" electrode of the 14-gap 704.4 MHz CH cavity, therefore, a higher $v_{water,max}$ would be necessary. However, as noted in the introduction, copper erosion-corrosion rates increase significantly when $v_{water,max}$ exceeds 1.2 m/s.

TABLE V. Comparison of ANSYS Discovery simulation results of the CH electrode with lotus-root-like cooling channels for the 14-gap 704.4 MHz CH cavity.

| Parameter | Case (a) | Case (b) | Case (c) |
|---|---|---|---|
| Material | pure copper | CuCr1Zr | CuCr1Zr |
| * Applied thermal conductivity $\kappa$ [W/(m·K)] | 397 | 310 (100%) | 263.5 (85%) |
| * Total thermal load $P_{loss,electrode}$ [kW] | 1.6 | 2.5 | 2.5 |
| * Outlet water pressure $p_{water,o}$ [bar] | 6 | 6 | 6 |
| * Inlet water temperature $T_{water,i}$ [°C] | 17 | 17 | 17 |
| * Inlet water flow rate $\dot{m}$ [kg/s] | 0.028 | 0.084 | 0.084 |
| # Outlet water temperature $T_{water,o}$ [°C] | 30.9 | 24.1 | 24.4 |
| # Max. water velocity $v_{water,max}$ [m/s] | 1.22 | 3.48 | 3.48 |
| # Max. surface temperature on the electrode $T_{surface,max}$ [°C] | 60.2 | 59.2 | 61.4 |

*: simulation conditions.
#: simulation results.

Beyond commonly used materials such as pure copper and 316L stainless steel, high-performance copper alloys such as CuCr1Zr and CuNi2SiCr are also being actively explored for MAM fabrication of accelerator components by the RACERS team at GSI [5]. TABLE II compares the key properties of these materials. Notably, these copper alloys, particularly CuCr1Zr, exhibit hardness comparable to 316L stainless steel while maintaining significantly higher thermal and electrical conductivity.

As shown in Case (b) in TABLE V and FIG. 14, a CuCr1Zr electrode can dissipate a maximum heat load of 2.5 kW at $v_{water,max} < 3.5$ m/s while maintaining $T_{surface,max} < 60°C$. The corresponding mass flow rate is 0.084 kg/s (approximately 5 L/min), which is consistent with the value applied in the thermal simulations for the MYRRHA CH cavities with stainless-steel electrodes. Even when assuming a reduced thermal conductivity (85% of the reference CuCr1Zr value) under otherwise similar conditions, repeated simulations showed that $T_{surface,max}$ increased only by 2.2°C, reaching 61.4°C. In short, such CuCr1Zr electrodes with lotus-root-like channels as well as the entire 14-gap 704.4 MHz CH cavity (including the end lids and the cavity wall, which exhibit lower power losses and simpler geometries) can be effectively cooled for both CW and pulsed operation modes.

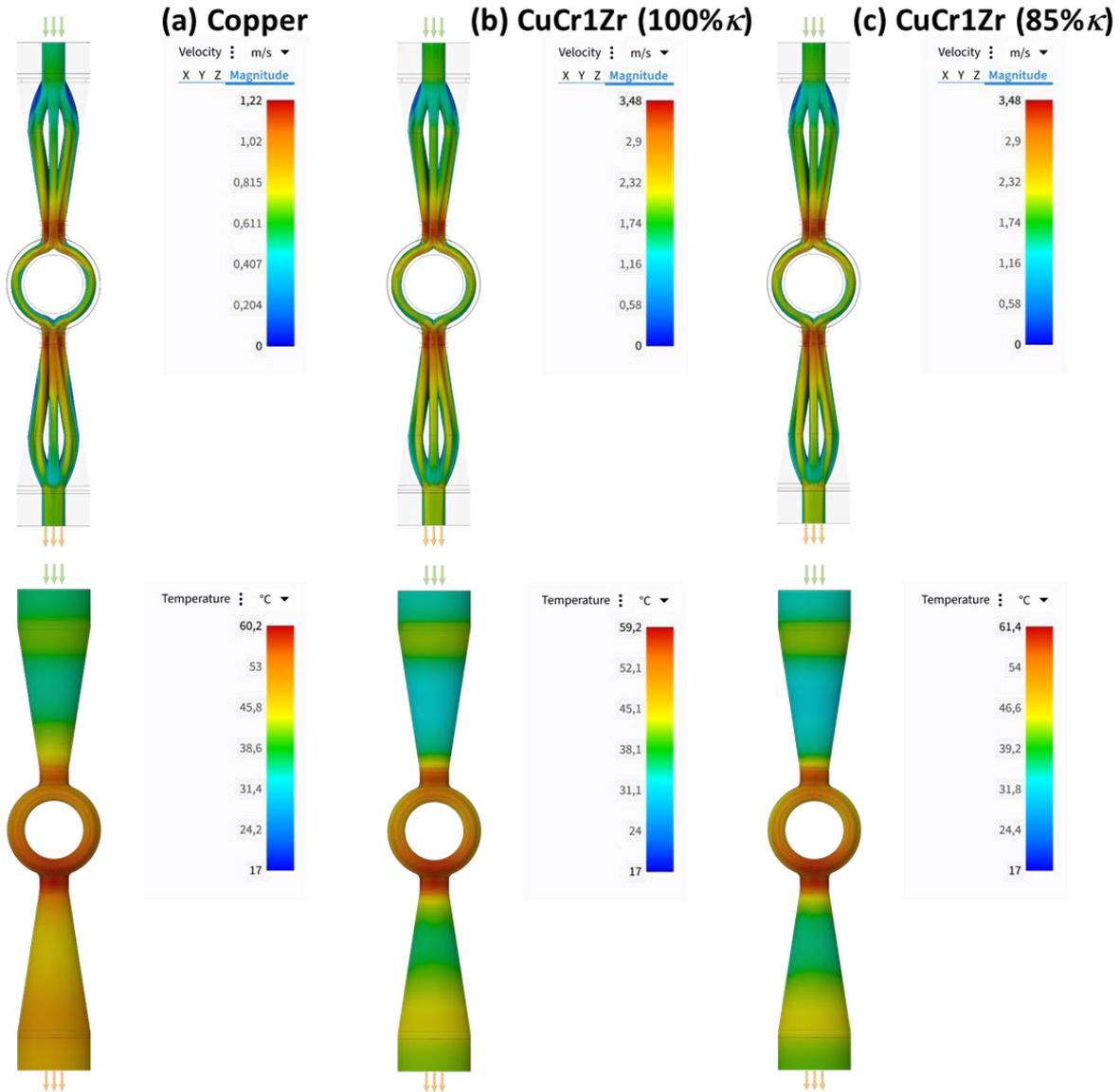

FIG. 14. Copper vs. CuCr1Zr electrodes with lotus-root-like cooling channels for the 14-gap 704.4 MHz CH cavity. ANSYS Discovery simulations of water-velocity (top) and surface-temperature (bottom) distributions targeting $T_{surface,max} \approx 60°C$. Arrows indicate the water-flow directions, and detailed results are summarized in TABLE V.

Based on the electrical conductivity of CuCr1Zr listed in Table II ($\sigma = 43$ MS/m, though later we will see that the measured $\sigma$ of a printed CuCr1Zr plate exceeds 48 MS/m), RF simulations show that the shunt impedance $Z_{eff}$ decreases from 45.1 MΩ/m to 41.1 MΩ/m. For the CW-design, copper plating can be applied to the inner surfaces of the cavity to mitigate this increase in power loss. For the Pulsed-design, the average power loss $P_{loss,electrode}$ of the "hottest" electrode is 2.3 kW, as mentioned in the previous section. The additional 9% increase in power loss raises the total $P_{\text{loss,electrode}}$ to approximately 2.5 kW, suggesting that copper plating may not even be necessary in this case.

Furthermore, the impact of frequency shifts due to thermally induced mechanical deformation was investigated. The linear thermal expansion can be expressed as:

$$\Delta L/L = \alpha\, \Delta T \tag{6}$$

where $\Delta L/L$ is the relative linear expansion in %, $\alpha$ is the linear expansion coefficient in K$^{-1}$ (°C$^{-1}$), and $\Delta T$ is the temperature change in K (°C). Consequently, a temperature rise from 17°C to 60°C leads to a linear expansion of 0.0731% for CuCr1Zr, e.g., the 20.3 mm cavity-wall thickness of the 14-gap 704.4 MHz CH cavity could expand by 0.015 mm. With large safety margins, CST simulations indicate that a uniform thermal expansion of the cavity's inner surface by 0.1 mm results in a frequency increase of approximately 0.5 MHz, whereas a 0.2 mm expansion leads to a shift of about 1.2 MHz. These frequency deviations remain well within the tuning range of the foreseen tuners.

For constructing H-mode drift tube linacs, CuCr1Zr is a kind of new material. Therefore, the electrical and thermal conductivity, surface quality, and vacuum compatibility of MAM-fabricated CuCr1Zr components must be carefully verified to ensure that they can meet the stringent requirements of accelerator applications. Currently, within the Stochastic Cooling Optimization (SCO) project [5], another MAM-focused initiative by the RACERS team at GSI, a CuCr1Zr-printed cooling plate, known as SCO1.0, has been developed. That R&D effort aims to optimize the water-cooling efficiency of the kickers in a 1–2 GHz stochastic cooling system for the FAIR project [5, 8].

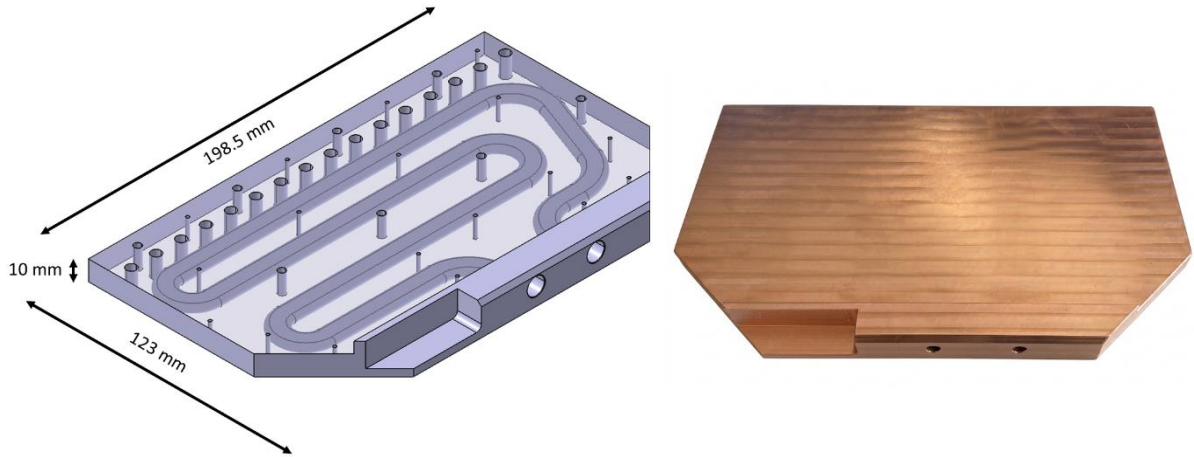

FIG. 15. The CuCr1Zr SCO1.0 cooling plate: mechanical design (left); the printed and post-processed part (right).

The CuCr1Zr cooling plate was fabricated through MAM and post-processed by the Fraunhofer Institute for Material and Beam Technology (IWS) in Dresden, Germany [32]. The electrical conductivity of the post-processed plate was measured at IWS using a Sigmascope 350 eddy current instrument (Helmut Fischer GmbH) equipped with an FS40 probe. More than ten measurements indicated that the electrical conductivity of the printed plate ranged from 83% to 86% IACS (International Annealed Copper Standard), which exceeded the reference value listed in TABLE II. Established by the International Electrotechnical Commission in 1913, 100% IACS corresponds to a conductivity of exactly 58 MS/m at 20°C, a value based on a specific high-purity annealed copper wire. Although the thermal conductivity was not directly measured, it can be estimated using the Wiedemann–Franz law. Since the dominant electronic component of thermal conductivity is linearly proportional to the electrical conductivity at a given temperature, it can be inferred that the printed SCO1.0 cooling plate should exhibit a thermal conductivity of approximately 310 W/(m·K).

For evaluating surface roughness, $R_a$ and $R_z$ are the two most critical parameters. While $R_a$ is defined as the arithmetic average of the absolute profile deviations over the entire evaluation length, $R_z$ represents the mean peak-to-valley height averaged across several consecutive sampling lengths. The surface quality of the post-processed CuCr1Zr cooling plate was measured at the GSI Technology Lab. Four reference points were selected (three on the front/back surfaces and one on the side surface), with measurements conducted in both horizontal and vertical directions. On the front and back surfaces, the measured $R_a$ values ranged from 0.342 to 0.840 μm, while $R_z$ values ranged from 2.315 to 3.634 μm. The side surface exhibited slightly higher roughness, with $R_a$ ranging from 0.744 to 1.377 μm and $R_z$ between 7.310 and 13.002 μm. In short, all these results are well within the ≤ ±20 μm tolerance typically required for RF structures.

Finally, vacuum performance and outgassing rates of the printed SCO1.0 cooling plate were characterized using a test chamber equipped with an ionization vacuum gauge and a mass spectrometer. After 1.5 hours of pumping, the vacuum pressure reached $1.7 \times 10^{-7}$ mbar, improving to $4.6 \times 10^{-8}$ mbar after 21 hours. These results confirm the component's vacuum compatibility for linacs, where $10^{-7}$ mbar is typically sufficient. Further measurements were performed under bake-outs at temperatures up to 140 °C, yielding the following results:
- Following the initial bake-out, the pressure reached $6.9 \times 10^{-9}$ mbar after 45 hours.
- After a second bake-out, the pressure further decreased to $3.8 \times 10^{-9}$ mbar after 69 hours.

Figure 16 displays the mass spectrum of residual gases for the printed SCO1.0 cooling plate. The primary residual gases identified were $H_2O$, $N_2$, CH–N+O, and $CO_2$, all exhibiting partial pressures below $2.0 \times 10^{-11}$ mbar. These residual gases most likely originate from the internal cooling channels, which did not undergo post-processing. In practical applications, however, these channels will be isolated from the vacuum environment, suggesting even better performance during operation.

In summary, CuCr1Zr-printed components demonstrate excellent suitability for linac applications from a vacuum perspective. For stochastic cooling applications (the primary objective of the SCO1.0 R&D), which demand ultra-high vacuum conditions, further investigations into performance optimization will be continued.

Therefore, all SCO1.0 cooling-plate R&D results confirm that CuCr1Zr is a feasible construction material for the 14-gap 704.4 MHz CH cavity and a promising candidate for future RF linacs.

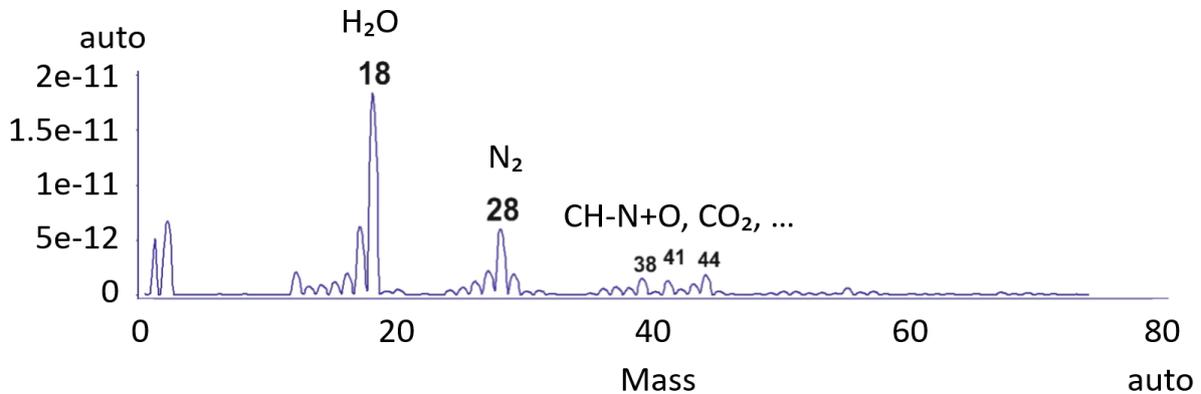

FIG. 16. The measured mass spectrum of the SCO1.0 cooling plate after a cumulative pumping time of 69 hours.

## V. CONCLUSIONS

To achieve efficient beam acceleration with high gradients is a primary objective in accelerator physics. For room-temperature structures, this pursuit is traditionally constrained by two critical challenges: RF sparking and thermal management. This study demonstrates that operating in the ultra-high frequency (UHF) range not only reduces the physical footprint but also significantly enhances sparking resistance. Our results indicate that the maximum surface electric field is no longer a limiting factor for the 14-gap 704.4 MHz CH cavity.

To overcome the cooling challenges inherent in the compact transverse dimensions of UHF structures, this study leverages the design freedom enabled by MAM. A unique cooling configuration featuring "lotus-root-like" channels, integrated with the high-strength, high-conductivity copper alloy CuCr1Zr, was successfully implemented. Comprehensive evaluations including beam dynamics, RF optimization, thermal analysis, and material characterization (surface quality, vacuum compatibility, and conductivity, …) validate the feasibility of this MAM-based approach. The 14-gap 704.4 MHz CH structure can achieve energy gain rates of 1.4–1.5 MeV/m (CW) and 4.6–4.8 MeV/m (pulsed), while maintaining a peak surface temperature of approximately 60°C. Future efforts will focus on optimizing the effective shunt impedance and cooling efficiency by further exploiting MAM's unique capabilities. Additionally, to go beyond the current accelerating-gradient limits, the development of a superconducting version of the 704.4 MHz CH structure (as well as other UHF structures) via MAM will be explored.

**ACKNOWLEDGEMENTS**

The author C. Z. thanks Dr. R. Tiede (IAP Frankfurt) for invaluable discussions on EQUUS design and LORASR code, Dr. K. Kümpel and S. Wagner (IAP Frankfurt) for useful information about the MYRRHA CH cavities, Dr. M. Eshraqi (European Spallation Source) and Dr. Y. Liu (Japan Proton Accelerator Research Complex) for their generous help and ~17 MeV particle distribution information from the ESS and J-PARC linacs.

The RACERS team acknowledges:
- Copper plating of the RACE cavity: Dr. R. Gebel (GSI), Dr. R. Stassen, Dr. F.-M. Esser, H. Schneider, B. Poschen, B. Cramer, W. Lesmeister, and S. Baier (Forschungszentrum Jülich).
- R&D of the SCO1.0 cooling plate: S. Gruber (Fraunhofer IWS Dresden) for MAM fabrication and post-processing; N. Kähne and M. Henke (GSI) for surface quality measurements; A. Bardonner (GSI) for vacuum tests.